\documentclass[manuscript]{acmart}

\setcopyright{none}
\AtBeginDocument{%
  }

\definecolor{lightyellow}{RGB}{255, 255, 204}
\definecolor{lightorange}{RGB}{255, 230, 204}
\definecolor{lightred}{RGB}{255, 204, 204}

\newcommand{\revise}[1]{\textcolor{black}{\textbf{}{#1}}}

\usepackage{booktabs}   
\usepackage{multirow}   
\usepackage{rotating}   
\usepackage{makecell}   
\usepackage{array}      
\usepackage{adjustbox}  
\usepackage{enumitem}
\usepackage{hyperref}
\usepackage{tabularx}
\usepackage{booktabs}
\usepackage{multirow}

\begin{document}

\title{The Governance of Intimacy: A Preliminary Policy Analysis of Romantic AI Platforms}


\author{Xiao Zhan}
\orcid{0000-0003-1755-0976}
\email{xzhan1@upv.es}
\affiliation{%
\institution{VRAIN, Universitat Politècnica de València \& University of Cambridge}
\city{Valencia}
    \country{Spain}
    \city{Cambridge}
    \country{UK}
}

\author{Yifan Xu}
\orcid{0000-0003-2303-1531}
\email{yifan.xu@manchester.ac.uk}
\affiliation{%
\institution{The University of Manchester}
\city{Manchester}
    \country{UK}
}

\author{Rongjun Ma}
\email{rma1@upv.es}
\orcid{0000-0001-7298-7762}
\affiliation{%
\institution{VRAIN, Universitat Politècnica de València \& Aalto University}
\city{Valencia}
    \country{Spain}
    \city{Espoo}
    \country{Finland}}

\author{Shijing He}
\orcid{0000-0003-3697-0706}
\email{shijing.he@kcl.ac.uk}
\affiliation{%
\institution{King’s College London}
\city{London}
\country{UK}}

\author{Jose Luis Martin-Navarro}
\email{jomarna6@upv.es}
\orcid{0000-0002-4503-4189}
\affiliation{%
\institution{VRAIN, Universitat Politècnica de València \& Aalto University}
\city{Valencia}
    \country{Spain}
    \city{Espoo}
    \country{Finland}}

\author{Jose Such}
\orcid{0000-0002-6041-178X}
\email{jose.such@csic.es}
\affiliation{%
\institution{INGENIO (CSIC-Universitat Politècnica de València)}
\city{Valencia}
\country{Spain}}
\renewcommand{\shortauthors}{Zhan et al.}

\begin{abstract}

Romantic AI platforms invite intimate emotional disclosure, yet their data governance practices remain underexamined. This preliminary study analyzes the Privacy Policies and Terms of Service of six Western and Chinese romantic AI platforms. We find that intimate disclosures are often positioned as reusable data assets, with broad permissions for storage, analysis, and model training. We identify default training appropriation, ownership reconstruction, and intimate-history assetization as key mechanisms structuring these practices, expanding platforms’ rights while shifting risk onto users. Our findings surface key governance challenges in romantic AI and are intended to provoke discussion and inform future empirical and design research on human–AI intimacy and its governance.

\end{abstract}



\maketitle

\section{Introduction}

Romantic AI companions are rapidly moving into the mainstream~\cite{statistic_1,statistic_2,statistic_3}, offering emotionally expressive, relationship-like interactions~\cite{marriage-children,marriage-children2,pataranutaporn2025my} that encourage users to disclose deeply personal thoughts, routines, and desires~\cite{ta2020user,skjuve2021my,ma2026privacy}. 
Unlike general-purpose AI systems, these platforms cultivate attachment, emotional dependence, and continuous self-disclosure~\cite{marriage-children,marriage-children2,pataranutaporn2025my,ho2025potential}, creating forms of vulnerability that exceed those found in ordinary chatbots. Over time, such interactions accumulate into intimate conversational histories that users may experience as private and relational.

Despite this heightened intimacy, little is known about how romantic AI platforms govern the sensitive data they elicit. Prior work has documented privacy discrepancies in individual platforms such as Replika~\cite{piispanen2024smoke}, 
but existing work has not examined romantic AI governance across multiple platforms or jurisdictions. Moreover, no work has investigated how the introduction of generative AI (GenAI) reshapes data governance, ownership, and safety responsibilities in emotionally intimate contexts. Finally, governance documents themselves have not been studied as relational infrastructures. As a result, we still lack clarity on how platforms legally frame intimacy, assign responsibility, and convert affective exchanges into model-training resources.

To address this gap, we conducted a qualitative policy analysis of six romantic AI governance, including three Western platforms and three non-Western (Chinese) platforms operating under distinct regulatory environments. Through qualitative analysis of their Privacy Policies (PPs) and Terms of Service (ToS), we investigate how platforms articulate data practices, define rights over GenAI-generated content, and claim user protections. Specifically, we pose the following 
research questions: 

\textbf{RQ1} (Data governance): \textit{How do romantic AI platforms describe their privacy practices regarding data collection, sharing, storage, deletion, and ownership?} 

\textbf{RQ2} (\revise{GenAI transparency}): \textit{How do romantic AI platforms \revise{disclose the details of their GenAI use, particularly regarding training data practices and output responsibility}? }

\textbf{RQ3} (User protection): \textit{What safeguards do romantic AI platforms claim to provide to keep their users safe?}


By answering these research questions, we make the following contributions: (1) This paper provides the first systematic examination of GenAI-related disclosures in romantic AI governance documents. (2) We uncover major transparency gaps and show that intimate conversations are treated as extractable resources, while responsibility for GenAI-related risks (e.g., hallucinated outputs, model training memorisation) is shifted onto users. (3)  Building on these findings, we argue that intimate data requires dedicated governance and call for redesigned consent mechanisms that account for emotional dependency and irreversible model training. (4) We position this work as a prequel for future empirical and design research, opening new conversations on intimate data governance, relational vulnerability, and consent in human–AI intimacy.


\section{\revise{Related Works}}
Early research on AI companionship highlights severe privacy and ethical challenges in platforms that mediate romantic or emotionally intimate relationships with artificial agents. Piispanen et al.~\cite{piispanen2024smoke} conducted a close reading of Replika’s Privacy Policy and compared it against public media reports, revealing that even GDPR compliant documents can obscure exploitative practices such as broad data collection, behavioral profiling, and emotional manipulation of vulnerable users. These findings echo broader concerns that AI companions operate under extreme information asymmetry, where service providers accumulate vast stores of intimate behavioral data that can be aggregated into user profiles~\cite{malki2024mhealth,shanmugarasa2025sok}. Journalistic and technical audits further expose weak privacy and security protections among AI-companion providers~\cite{ragab2024trust}, showing that data about daily routines, sexuality, or health-related experiences can be inferred from conversational logs and potentially accessed by third parties.

Beyond system-level data practices, recent studies explore how users themselves navigate disclosure with AI partners. Wang et al.~\cite{wang2025my} identify two dominant orientations toward self-disclosure: one group views openness with AI as natural and beneficial given its perceived emotional support, while another expresses apprehension about surveillance and misuse of sensitive data. Similarly, \citet{djufril2025love} find that users who report stronger emotional attachment to their AI partners tend to share more personal information than they would with human partners, though others remain selective depending on topic or context. Recent research~\cite{ma2026privacy} further shows that as emotional attachment deepens, users often deprioritize privacy concerns and disclose more intimate information, perceiving AI partners as less risky than human partners while remaining wary of platform-level surveillance and data retention practices. Parallel qualitative analyses of user forums and social media communities~\cite{pataranutaporn2025my} suggest that users often normalize or even romanticize intimate data exchange, treating disclosure as a sign of connection rather than a privacy risk.

At a broader ethical level, Ho et al.~\cite{ho2025potential} synthesize the potential and pitfalls of romantic AI systems, identifying recurrent risks around data misuse, user manipulation, and emotional dependency. However, as Dewitte~\cite{dewitte2024better} notes, few of these studies examine the policy layer that ostensibly governs such risks—namely the Privacy Policies and Terms of Service through which platforms legally frame user consent. Recent theoretical and audit-based research therefore calls for integrated policy analyses that connect declared privacy practices with the technical and emotional realities of human–AI intimacy~\cite{kouros2024digital,harmfulai2025traits,mireshghallah2024wildchat}.

\section{Methodology}

\subsection{Platform and Document Selection} \label{method_platform_selection}



To capture a diverse range of governance approaches, we selected six romantic AI companionship platforms based on three criteria: (i) Popularity, as indicated by high download rankings in their respective regional markets; (ii) LLM-based; and (iii) Regional Representation, with three platforms from Western markets (\href{https://grok.com/ani}{Grok Ani}, \href{https://nomi.ai/}{Nomi.AI}, and \href{https://replika.com/}{Replika}) and three from non-Western markets, specifically Chinese domestic platforms (\href{https://maoxiangai.com/}{Maoxiang}, \href{https://www.zhumengdao.com/}{Zhumengdao}, and \href{https://www.xingyeai.com/}{Xingye}). \footnote{Grok Ani: \url{https://grok.com/ani}, \\ Nomi: \url{https://nomi.ai/}, \\Replica: \url{https://replika.com/}, \\ Maoxiang: \url{https://maoxiangai.com/}, \\ Zhumengdao: \url{https://www.zhumengdao.com/}, \\ Xingye: \url{https://www.xingyeai.com/}}.
We collected and analysed both the PPs and the ToS for each platform. The PPs primarily function as regulatory disclosures that describe a platform’s data practices, whereas the ToS operates as a broader contractual agreement that defines the rules governing platform use. Examining the ToS is particularly important for romantic AI platforms because key GenAI governance provisions, including policies on ownership of AI-generated content and statements about model hallucinations, are often located in the ToS rather than in the PPs. We therefore analyze both documents, following prior work that emphasizes the value of studying them together, since a joint analysis can reveal contradictions, redundancies, and misalignments in how platforms present their data and governance practices across these documents~\cite{palka2021big,soneji2024demystifying,andow2019policylint,okoyomon2019ridiculousness}.

Examining both documents also helps reveal potential inconsistencies or forms of \textit{``policy decoupling''}. For example, promises of user data control in the PPs may be undermined by broad and perpetual licensing rights in the ToS, or the PPs may even contradict itself~\cite{andow2019policylint, okoyomon2019ridiculousness}.

\subsection{Qualitative Coding Methodology}



We employed a hybrid qualitative content analysis combining inductive theme development with deductive coding. Drawing from prior research~\cite{he2026investigating,javed2021privacy, malki2024exploring}, we applied major regulatory frameworks---GDPR~\cite{GDPR} and the CCPA~\cite{CCPA} for the three Western platforms; and PIPL~\cite{PIPL} for platforms operating in the Chinese market---as our 
high-level conceptual guides for drafting the initial codebook. These frameworks informed which categories (e.g., data types, processing purposes, retention, user rights, and safety obligations) should be included, but they were not used as coding schemas. Guided by these, we developed themes and their associated codes for RQ1, including themes such as \textit{``Data Types Collected'', ``Data Collection Purpose'', ``Data Retention \& Determination''}, and \textit{``Data Sharing Recipients''}. We also generated themes relevant to RQ3, such as \textit{``Technical Security Measures''}. For RQ2, the authors met to discuss and inductively develop themes specific to GenAI–related risks, including \textit{``Training Data Sources'', ``Opt-Out Mechanisms for Model Training''}, and \textit{``Accuracy or Hallucination Disclaimers''}. After establishing the initial codebook, two authors independently coded all documents, while the remaining authors cross-checked segments and joined discussions of ambiguous cases. Throughout the coding process, the research team refined the codebook through regular meetings, merging overlapping codes and clarifying category definitions. We did not calculate inter-rater reliability (IRR), as our interpretivist, consensus-based approach prioritized iterative discussion in codebook development~\cite{ortloff2023different}. We ensured analytic credibility through collaborative reconciliation and systematic cross-checks, rather than relying on numeric agreement metrics~\cite{mcdonald2019reliability}. 

\section{Policy Analysis Findings}
\label{sec:findings}

Before presenting the RQ findings, we compiled a descriptive overview of the PPs and ToS (e.g., the exact document version used, accessibility, update cycles). 


\subsection{Overview}
\revise{All formal governance documents are accessible through the platform’s user interface, both on web-based platforms (via the settings or account menus) and within mobile applications (typically under ``Account'', ``Privacy'', or ``Legal'' sections). While this accessibility is a common baseline, the level of detail, update frequency, and clarity of communication vary notably across platforms.}

\revise{Regarding how these platforms handle policy changes. Across the six platforms (see table \ref{tab:policy_updates}), 
all PPs and 5 ToS documents clearly display their effective dates. Only Grok's platform provides access to previous terms of service versions for comparison. Additionally, Xingye and Zhumengdao explicitly state that they will notify users about major changes; notifications tend to appear as in-app messages or banners rather than personalised emails, offering a slightly more proactive stance. However, several platforms default to the assumption that continued use of the service constitutes user consent to the updated terms. For example, Grok’s ToS explicitly states that for global users, notification of changes is satisfied simply by updating the ``Effective Date” at the top of the document, with no affirmative obligation to email or directly inform users.}

\revise{Most platforms do not offer access to previous versions of their policies. Only Grok maintains a publicly accessible archive of its ToS, making it uniquely traceable for users or researchers interested in how its governance practices have changed. In terms of how current their policies are, three platforms Grok, Zhumengdao, and Xingye have revised their governance documents within the last year. Grok, for instance, last updated its policy within the past six months, reflecting its ongoing adaptation to a rapidly changing AI environment. Nomi, however, has not published substantial revisions since 2023. }

\begin{table}[htbp]
    \centering
    \scriptsize
    \setlength{\tabcolsep}{3pt}
    \renewcommand{\arraystretch}{1.3}
    \caption{Privacy Policy and Terms of Service Updates Summary}
    \label{tab:policy_updates}
    \resizebox{0.6\linewidth}{!}{
    \begin{tabular}{l | c | c | c | c}
        \toprule
        \textbf{Platform} & 
        \textbf{Version Analysed} &
        \textbf{Latest Version} &
        \textbf{Contacts} &
        \textbf{Version History} \\
        \midrule

        \textbf{Grok (PP)} & 
        10/07/2025 & 
        10/07/2025 & 
        Provided & --
        \\

        \textbf{Grok (PP–EU)} & 
        24/04/2025 & 
        24/04/2025 & 
        Provided &--
        \\

        \textbf{Grok (ToS)} & 
        04/11/2025 & 
       04/11/2025& 
        Provided &Available
        \\
    \midrule

        \textbf{Nomi (PP)} & 
        14/04/2023 & 
        14/04/2023 & 
        Provided &--
        \\

        \textbf{Nomi (ToS)} & 
        -- & 
        -- & 
        -- &--
        \\
    \midrule

        \textbf{Replika (PP)} & 
        01/11/2025 & 
        01/11/2025 & 
        Provided &--
        \\

        \textbf{Replika (ToS)} & 
       07/02/2023 & 
       07/02/2023
         & 
        Provided &--
        \\
    \midrule

        \textbf{Maoxiang (PP)} & 
        02/12/2025 & 
        02/12/2025-- & 
        Provided &--
        \\

        \textbf{Maoxiang (ToS)} & 
       04/11/2025 & 
       04/11/2025  & 
        -- &--
        \\
    \midrule

        \textbf{Zhumengdao (PP)} & 
        18/06/2025 & 
         18/06/2025 & 
        Provided &-
        \\

        \textbf{Zhumengdao (ToS)} & 
        18/06/2025 & 
        18/06/2025 & 
        Provided &--
        \\
    \midrule

        \textbf{XingYe (PP)} & 
        17/11/2025 & 
        17/11/2025& 
        Provided &--
        \\

        \textbf{XingYe (ToS)} & 
        01/09/2025 & 
        01/09/2025 & 
        Provided &--
        \\
        
        \bottomrule
    \end{tabular}}
\end{table}

\begin{table*}[t]
    \centering
    \caption{Presence of Data Type Collections in PP and ToS of Romantic AI Platforms}
    \label{tab:data_types}
    \vspace{-1.5ex}
    \resizebox{\linewidth}{!}{
        \begin{tabular}{l | *{10}{cc}}
            \toprule

            \textbf{} 
            & \multicolumn{2}{c}{Account}
            & \multicolumn{2}{c}{Payment}
            & \multicolumn{2}{c}{Communication}
            & \multicolumn{2}{c}{Interests \& Preferences}
            & \multicolumn{2}{c}{Usage }
            & \multicolumn{2}{c}{Feedback }
            & \multicolumn{2}{c}{Social Media }
            & \multicolumn{2}{c}{Technical }
            & \multicolumn{2}{c}{Publicly Available}
            & \multicolumn{2}{c}{Face \& Head Movement} \\
            
            \cmidrule(lr){2-3}
            \cmidrule(lr){4-5}
            \cmidrule(lr){6-7}
            \cmidrule(lr){8-9}
            \cmidrule(lr){10-11}
            \cmidrule(lr){12-13}
            \cmidrule(lr){14-15}
            \cmidrule(lr){16-17}
            \cmidrule(lr){18-19}
            \cmidrule(lr){20-21}

            & PP & ToS & PP & ToS & PP & ToS & PP & ToS & PP & ToS
            & PP & ToS & PP & ToS & PP & ToS & PP & ToS & PP & ToS \\
            
            \midrule

            \textbf{Grok}
            & \checkmark & \checkmark
            & \checkmark & \checkmark
            & \checkmark & \checkmark
            & --- & ---
            & \checkmark & \checkmark
            & \checkmark & \checkmark
            & \checkmark & \checkmark
            & \checkmark & \checkmark
            & \checkmark & \checkmark
            & --- & --- \\
            
            \textbf{Nomi}
            & \checkmark & ---
            & \checkmark & ---
            & \checkmark & ---
            & --- & ---
            & \checkmark & ---
            & --- & ---
            & --- & ---
            & \checkmark & ---
            & --- & ---
            & --- & --- \\
            
            \textbf{Replika}
            & \checkmark & \checkmark
            & \checkmark & ---
            & \checkmark & ---
            & \checkmark & ---
            & \checkmark & ---
            & --- & ---
            & \checkmark & \checkmark
            & \checkmark & ---
            & --- & ---
            & \checkmark & --- \\
            
            \textbf{Maoxiang}
            & \checkmark & \checkmark
            & \checkmark & ---
            & \checkmark & \checkmark
            & --- & ---
            & --- & ---
            & \checkmark & ---
            & \checkmark & ---
            & \checkmark & ---
            & \checkmark & ---
            & \checkmark & --- \\
            
            \textbf{Zhumengdao}
            & \checkmark & \checkmark
            & \checkmark & ---
            & \checkmark & \checkmark
            & \checkmark & ---
            & \checkmark & ---
            & \checkmark & \checkmark
            & \checkmark & ---
            & \checkmark & ---
            & \checkmark & ---
            & --- & --- \\
            
            \textbf{Xingye}
            & \checkmark & \checkmark
            & --- & ---
            & \checkmark & \checkmark
            & --- & \checkmark
            & --- & ---
            & \checkmark & ---
            & \checkmark & ---
            & \checkmark & ---
            & \checkmark & \checkmark
            & --- & --- \\
            
            \bottomrule
        \end{tabular}
    }
    
    \raggedright \footnotesize \textit{Note: \checkmark means Explicitly mentioned as collected; ---denotes Not mentioned.}
    \vspace{-8pt}
\end{table*}

\subsection{Data Governance (RQ1)}
\label{sec:data-governance}




\subsubsection{Data Types Collected and Purpose}
\label{sec:datatype-collection}

Tab. \ref{tab:data_types} summarises the data types that platforms explicitly state they collect. Additional details on the purposes of data collection and the recipients with whom these data may be shared are provided in Appendix \ref{appendix:data-practices} (see Tab. \ref{tab:account_data}-\ref{tab:public_data}).

Across the six platforms, disclosures about data collection differ sharply in granularity and completeness, meaning users face fundamentally unequal levels of knowability about what is collected and why. Grok and Replika offer the most structured disclosures, providing category-level descriptions and linking data types to specific purposes. By contrast, Nomi and several Chinese platforms disclose far less detail. For example, Nomi PP states only that it collects information users \textit{``directly provide to us''} or \textit{``that is generated when they interact with the Services''}, without specifying what these categories include. Document inconsistency further undermines transparency: some data types appear in the PP but not the ToS (Nomi, Zhumengdao), while use purposes appear in one document but not the other (Xingye, Maoxiang), requiring users to reconcile fragmented and incomplete statements.

Despite these disparities, all platforms name the same three core data types somewhere in their documents: account-, communication-, and technical data. Several also reserve rights to ingest \textit{``publicly available information''} or feature-dependent data such as payment details, extending profiling beyond the romantic interface. Stated purposes similarly range from specific to highly open-ended. All platforms justify processing communication data for service provision or \textit{``security, integrity, and legal compliance''}, yet most simultaneously authorise broad secondary uses such as improvement, and analytics. Some (e.g., Replika and Maoxiang) explicitly include marketing as a purpose; by contrast, Grok explicitly excludes it, and a majority omit any mention entirely. This leaves 
users unable to determine whether marketing occurs or is simply undisclosed. These selective disclosures create asymmetric transparency: platforms name data types but withhold clarity about which purposes apply and how intimate data flows through internal processes.

\subsubsection{Data Sharing and Third-Party Access}
\label{sec:data-sharing}

All platforms acknowledge that user data may be shared with third parties, though transparency varies substantially. Grok and Replika offer the clearest breakdowns in their PPs, identifying categories such as service providers, affiliates, and law-enforcement authorities.
Alarmingly, Grok (ToS) grants them a license to share user content for \textit{``any purpose''}, directly conflicting with its PPs. Replika (PP) similarly distinguishes between vendors supporting operations and entities receiving data for analytics or research. However, other platforms adopt less granular approaches. Nomi provides only a high-level statement that information may be shared as needed to provide \textit{``core functionality''}, without naming specific recipients. Chinese platforms generally list typical categories such as affiliated companies, service providers, and government or regulatory bodies. Despite these disclosures, no platform offers detailed criteria for when sharing is triggered, how frequently access occurs, or what technical controls limit third-party processing. As a result, while the existence of third-party access pathways is clear, the scope and conditions of such access remain broadly defined.

\subsubsection{Retention \& Deletion}
\label{sec:retention}
Across the six platforms, retention and deletion policies share a common structure but vary significantly in specificity. Most rely on broad clauses such as retaining data \textit{``for as long as necessary''} or \textit{``as required by law''}. Nomi is the only service to promise immediate erasure, stating that data will be \textit{``immediately deleted and cannot be recovered'' (Nomi ToS)}. Others provide more conditional timelines: Grok (PP) describes deletion requests entering a processing queue, while Maoxiang (PP) and Zhumengdao (PP) cite statutory log-retention requirements applicable in their jurisdictions. Replika (PP) offers detailed retention periods for different data categories but, unlike Grok, does not specify an operational deletion window. Across all platforms, this user-oriented framing coexists with broad ToS provisions allowing providers to remove content or terminate accounts at their discretion. As a result, formal deletion rights are paired with expansive platform authority over data removal and account shutdown.




\subsubsection{Ownership - IP Rights and Human Review}
\label{sec:ownership}

Across all six platforms, ownership language signals user control but is substantially limited by broad licensing and review rights. Western platforms allow users to retain copyright while simultaneously claiming sweeping licenses that undermine exclusivity, for example, Replika (ToS) requires a \textit{``perpetual, irrevocable, and sublicensable''} license, and Nomi (ToS) permits the use of user content in \textit{``financing, sale, [or] transfer''} of the service, effectively treating intimate chat histories as corporate assets. 
Chinese platforms likewise deploy an ``acknowledge-and-appropriate'' structure, with Xingye (ToS) further coupling its license with stringent user liability clauses. Regardless of these formal claims, all Western platforms reserve rights for \textit{``authorised personnel''} to access or review content, such as Grok’s review for \textit{``improving product features''} or investigating misuse (Grok PP). These provisions make clear that ownership---whether nominally held by users or claimed by platforms---does not prevent internal access for moderation, safety, or development. 

\subsection{Disclose the Use of GenAIs (RQ2)}
\label{sec:finding-genaiuse}


We examined how platforms disclose the specific mechanics of their LLMs, specifically regarding training data sources and liability for GenAI-generated outputs.

\subsubsection{Training Data Disclosure \& Use}
\label{sec:training-data}
Only Grok provides any information about pre-training sources, noting that its models use \textit{``publicly available information on the internet (Grok PP)''}. Other platforms avoid external data provenance and disclose only that user-generated content may be used to \textit{``train''}, \textit{``optimise''}, or \textit{``improve''} their systems (Replika PP; Nomi PP; Maoxiang PP; Xingye PP; Zhumengdao PP).

\subsubsection{Opt-Out Mechanisms}
\label{sec:opt-out}

Only three platforms offer any way to refuse model-training use. Grok (PP) allows users to disable training use directly through a settings toggle, Zhumengdao (PP) recognises a right to object but requires users to \textit{``contact us […] to request the withdrawal''}, making the process more labour-intensive. Maoxiang (PP) offers both opt-out approaches. All other platforms that state they use user content for training provide no opt-out mechanism, meaning user interactions are treated as default training inputs with no procedural control. However, a fundamental contradiction exists: the perpetual usage rights secured in these platforms' IP clauses (\S\ref{sec:ownership}) nullify the very purpose of their offered opt-out mechanisms.


\subsubsection{Disclaimers \& Allocation of Responsibility}
\label{sec:disclaimer}
Four platforms (Grok, Maoxiang, Zhumengdao, and Xingye) explicitly limit responsibility for AI reliability, though with different levels of specificity. Zhumengdao (PP) issues the strongest warning, requiring users to \textit{``independently verify''} outputs, \textit{``especially regarding numbers, time, and factual descriptions''}, and cautioning that accuracy cannot be \textit{``100 percent guaranteed''}. Maoxiang (PP) likewise states that outputs are “for reference only” and that users bear all consequences arising from reliance on their \textit{``authenticity, accuracy, or reliability''}, while Grok and Xingye use broader warranty disclaimers in their ToS that emphasise the AI service may be interrupted, erroneous, or fail to meet expectations and that errors need not be fully corrected. Across these documents, hallucination and other failures are framed as risks that users, rather than platforms, must ultimately absorb.

\subsubsection{Mandatory AI Labelling}

All three Chinese platforms commit to labelling AI-generated content in accordance with China’s 2023 AIGC Interim Measures~\cite{AILabel}. Their terms state that providers may add \textit{``labels''} or watermarks to outputs and that users must clearly mark AI-generated content when sharing it and may not remove such labels (e.g., Xingye ToS; Maoxiang ToS; Zhumengdao ToS). None of the Western platforms mentions comparable requirements, underscoring a regulatory divide in how transparency obligations are defined across regions.

\subsection{Platform-Claimed Safeguards (RQ3)}


\subsubsection{Age Safety \& Protection of Minors}
\label{sec:age-safety}


All six platforms formally claim to protect minors, but the strength and specificity of age-related safeguards vary considerably. Chinese platforms consistently frame minor protection as a platform-led governance responsibility, and they employ real-name verification to identify suspected underage users and apply mandatory restrictions accordingly. Maoxiang implements the most stringent regime: once identified as a minor, an account enters a restricted mode that \textit{``cannot like, create, comment, share, top-up, or consume''}, and real-name verified minors \textit{``cannot exit minor mode until they turn 18''}. In contrast, 
English-language platforms offer only minimal, declarative age safeguards. Replika and Nomi simply state that their services are for adults and may delete underage accounts, without describing any detection or enforcement. Grok allows users as young as 13 and openly warns that its outputs may include \textit{``sexual situations, violence, [and] crude humour''}, yet provides no technical restrictions or minor-protection mode beyond a basic reporting channel.

\subsubsection{Technical Security Measures}
\label{sec:technical-security-measures}

For security measures, Grok provides the most minimal commitments, its PP states only that they \textit{``implements commercially reasonable technical, administrative, and organisational measures''}, without offering any concrete mechanisms or explanations. Nomi likewise discloses almost no technical safeguards; its ToS even caution that data transmissions \textit{``may be unencrypted''}, making it the only platform to openly acknowledge the possibility of plaintext transfer. By contrast, the other platforms describe more concrete, system-level safeguards. Across their privacy policies, they collectively reference measures such as \textit{``SSL encryption, secure servers with firewalls, and role-based access control (Replika PP)''}, \textit{``encrypted storage, access-permission controls, and breach-notification mechanisms (Xingye PP)''}, \textit{``industry-standard encryption and independent encrypted storage of sensitive data (Maoxiang PP)''}, and \textit{``encrypted storage and transmission, strict access controls, and incident-response procedures (Zhumengdao PP)''}. Together, these disclosures present a notably more specific and multi-layered security posture, in sharp contrast to the minimal, generic assurances offered by Grok and Nomi.

\subsubsection{Content Safety \& Governance}
\label{sec:content-safety}
Across all six platforms, providers uniformly claim that they will intervene against unlawful or inappropriate content. Except for Nomi, every platform states that, upon detecting a violation, it may delete or block content, suspend or terminate accounts or services, and cooperate with authorities, including reporting severe cases. For example, Zhumengdao (PP) explicitly notes that it may \textit{``report relevant information to competent authorities in accordance with the law''} and Grok (ToS) states that they will \textit{``cooperate with law enforcement''}. What differs is the level of specificity: Maoxiang is the only platform that names an actual review pathway, noting that content may be processed by \textit{``third-party content review providers''}, whereas the others simply assert that they conduct review without describing how it operates. Nomi offers the least detailed regime. Its ToS merely state that it may remove \textit{``unlawful, defamatory, harassing, abusive, or otherwise objectionable''} content and terminate accounts, without further elaboration on moderation processes or enforcement mechanisms.

\section{Discussion}
\label{sec:discussion}




Building on our findings (§\ref{sec:data-governance}), platforms do not treat intimacy as a distinct category of sensitive data; instead, romantic disclosures are processed through the same pipelines as routine technical or account data, enabling broad reuse and retention with minimal constraints. Intimacy is further endangered by opaque data practices and, as our findings (\S\ref{sec:data-sharing},\S\ref{sec:ownership}) show, the commercial extraction of the user data far beyond their reasonable expectations. This reveals a deeper governance problem: platform policies systematically strip intimacy of its relational significance, recasting disclosures as assets rather than components of a private relationship. Our findings identify three mechanisms that make such commercial extraction possible (\S\ref{sec:ownership},\S\ref{sec:training-data}, \S\ref{sec:opt-out}). Default training appropriation renders all romantic interactions available for model optimization unless users actively object -- an option most platforms do not provide. Ownership reconstruction operates through expansive licenses or outright claims of corporate ownership, ensuring platforms retain downstream rights over affective histories regardless of users’ expectations of confidentiality. Intimate-history assetization, most visible in Nomi’s allowance of content use during financing or sale, transforms the accumulated traces of emotionally charged relationships into transferable corporate property. Together, these mechanisms illustrate how platforms convert relational exchanges into durable computational and economic resources.


Even where platforms nominally grant users ownership or control, intimate conversations lack meaningful privacy protections. Nomi’s ToS even acknowledges that data transmissions \textit{``may be unencrypted''}, exposing disclosures to interception (§\ref{sec:technical-security-measures}). Western platforms reserve broad rights for \textit{``authorised personnel''} to access user content and Maoxiang delegates review to third-party contractors (§\ref{sec:ownership}). These practices reveal that user ownership is largely symbolic: operational control resides with platform employees and external reviewers. The premise of a private, one-to-one relationship with an AI partner is thus incompatible with infrastructures that treat intimacy as content to be inspected, moderated, and repurposed.


Our preliminary analysis identifies two structural barriers limiting meaningful user control and raising key governance challenges. First, a temporal mismatch: consent is obtained at registration, long before emotional attachment forms, even though attachment reliably increases the depth \revise{and intimacy of} disclosure~\cite{ta2020user,skjuve2021my}. Users often misunderstand or overlook policy terms \revise{at the point of registration}~\cite{obar2020biggest, cranor2020effectively, nouwens2020dark}, and \revise{as emotional bonds deepen, they may disclose increasingly sensitive information~\cite{ma2026privacy}. Yet the consent governing these disclosures remains fixed, granted before users experience the system's relational dynamics.}
\revise{This temporal mismatch is further reinforced by a second structural constraint}: irreversibility. Even where platforms offer opt-out mechanisms, these typically operate only  prospectively; intimate disclosures already incorporated into model training cannot meaningfully be withdrawn (§\ref{sec:opt-out}). At the same time, liability disclaimers shift responsibility for GenAI-generated harm to users (§\ref{sec:disclaimer}). This creates a structural imbalance in which platforms benefit from progressive emotional vulnerability while users retain neither the ability to retract data nor recourse against downstream uses.

\paragraph{Limitations and Future Work.} \revise{As a preliminary study, this work has several limitations that warrant further development. First, future research should examine a broader range of romantic AI platforms, expanding not only the sample size but also the diversity of regional contexts in which these systems operate. Second, some of the governance mechanisms identified here, such as broad licensing clauses and liability disclaimers, also appear in general-purpose AI services and other digital applications (e.g., smart home platforms). Our analysis does not systematically compare romantic AI governance with that of non-romantic AI systems, as our focus was on understanding how governance operates within intimacy-oriented contexts. Future research could therefore considering conducting cross-domain policy comparisons to determine whether romantic AI introduces substantively distinct governance dynamics or instead amplifies existing platform practices through relational dependency. Third, }
future work should also explore how to better support users' understanding of governance risks and potential ``policy traps'' through clearer and more contextualised policy cues. Fifth, empirical research is needed to examine how users interpret and negotiate these governance conditions in practice, while traceability studies should compare platforms’ stated policies with their actual data practices. 
Last, \revise{there remains an interpretive question concerns how to understand the finding that platforms abstract relational interactions into technical and economic assets. One explanation is that companies view these systems primarily as technical tools and therefore apply standard governance models without fully addressing their relational implications. Yet, intentional or not, this framing overlooks the distinctive characteristics of intimate relationships. Because these systems are designed to cultivate emotional dependency, governance approaches that fail to recognise this specificity warrant critical scrutiny. Future work will be necessary to examine this interpretive tension with additional empirical evidence, particularly by investigating how designers understand the relational implications of the systems they build.}


\section{Conclusion}

This study examined Privacy Policies and Terms of Service across six romantic AI platforms to assess transparency and risk distribution. Our findings reveal platforms designed for emotional attachment while contractually treating intimate disclosures as commercial assets subject to training, transfer, and corporate control. Across the platforms examined, three recurring mechanisms enable this extraction: default training appropriation, ownership reconstruction, and intimate history assetization, sustained by consent obtained before dependency forms and \revise{rendered effectively irreversible after integration. }
We also recommend future work examining user perceptions, auditing the policy-practice gap, and co-designing consent mechanisms accounting for progressive dependency. Recognising relational asymmetry as a basis for data protection warrants serious consideration in efforts to prevent the commodification of human vulnerability.


\begin{acks}

We thank the anonymous reviewers for their constructive feedback, which has helped us improve this work at its early stage. This research was supported by the INCIBE's strategic SPRINT (Seguridad y Privacidad en Sistemas con Inteligencia Artificial) C063/23 project with funds from the EU-NextGenerationEU through the Spanish government's Plan de Recuperación, Transformación y Resiliencia, by the Generalitat Valenciana under grant CIPROM/2023/23, and grant PID2023-151536OB-100 funded by MICIU/AEI/10.13039/501100011033 and by ERDF/EU.
\end{acks}

\bibliographystyle{ACM-Reference-Format}
\bibliography{sample-base}

\newpage

\appendix

\section{Appendix}

\subsection{Data Practices Details}
\label{appendix:data-practices}

In this section, we present detailed definitions of the datatypes used, along with tables analysing their collection, usage, and disclosure.

\begin{itemize}
    \item Account Data (Table \ref{tab:account_data}): Information provided by the user when registering for, logging into, or maintaining an account on a romantic AI platform, such as name, phone number, email address, date of birth, and login credentials.
    \item Payment Data (Table~\ref{tab:payment_data}): Information related to financial transactions on the platform, including payment method details,  subscription information, and purchase history, whether processed directly by the platform or via third-party payment services.
    \item Communication Data (Table~\ref{tab:communication_data}): Content generated through interactions between the user and the AI, including user-provided inputs (e.g., text messages, voice recordings, images, or uploaded files) and AI-generated outputs (e.g., responses, dialogue, or role-play content), as well as the resulting conversation history.
    \item Interests \& Preferences Data (Table~\ref{tab:interest_data}): Data reflecting a user’s likes, dislikes, chosen conversation topics, interaction styles, or usage habits, including explicit selections and inferred preferences.
    \item Usage Data (Table~\ref{tab:usage_data}): Records describing how a user interacts with the platform, such as clicks, page views, navigation paths, session duration, feature usage, and other activity logs.
    \item Feedback Data (Table~\ref{tab:feedback_data}): Information provided by the user about their experience with the service, such as ratings, reviews, survey responses, bug reports, or improvement suggestions.
    \item Social Media Data (Table~\ref{tab:socialmedia_data}): Information obtained through a user’s connection to or interaction with external social media platforms, such as profile information, usernames, social connections, or content shared via linked accounts.
    \item Technical \& Location Data (Table~\ref{tab:technical_data}): Automatically collected device- and network-related information, such as IP address, device identifiers, operating system, browser type, application version, network information, and approximate location data.
    \item Publicly Available Data (Table~\ref{tab:public_data}): Information that is publicly accessible or obtained from public sources, such as publicly available online content or profiles.

\end{itemize}

\begin{table*}[htbp]
    \centering
    \scriptsize 
    \setlength{\tabcolsep}{2pt} 
    \renewcommand{\arraystretch}{1.3} 

    \caption{Account Data Collection, Usage, and Disclosure Analysis}
    \label{tab:account_data}
    \resizebox{\linewidth}{!}{
\begin{tabular}{l | c  | p{1cm} | *{7}{c} | *{7}{c} }
        \toprule
        \multirow{3}{*}{\textbf{APP}} 
        & \multicolumn{16}{c}{\textbf{Account Data}} \\
        \cmidrule(l){2-17}
        
        & \multirow{2}{*}{\rotatebox{90}{\textbf{Collect?}}} 
        & \multirow{2}{*}{\thead{Collect \\ Method} } 
        & \multicolumn{7}{c|}{\textbf{Use / Process}} 
        & \multicolumn{7}{c}{\textbf{Disclose / Share}} \\
        \cmidrule(lr){4-10} \cmidrule(l){11-17}

        & &  
        & \thead{Maintain \\ Service} 
        & \thead{User \\ Support} 
        & \thead{Improve / \\ Research} 
        & \thead{Communic- \\ ate} 
        & \thead{Security \\ Integrity} 
        & \thead{Legal \\ Purposes} 
        & \thead{Marketing} 
        & \thead{Service \\ Providers} 
        & \thead{Business \\ Transfers} 
        & \thead{Legal \\ Authorities} 
        & \thead{Related \\ Companies} 
        & \thead{Advertisers} 
        & \thead{Other 3rd \\ Parties} 
        & \thead{Public} \\
        \midrule

        \textbf{Grok (PP)} & 
        Yes & 
        
        \raggedright Direct  & 
        Yes & Yes & Yes & Yes & Yes & Yes & No & 
        Yes & Yes & Yes & Yes & No & Yes & -- \\ 
        \textbf{Grok (PP-EU)} & 
        Yes & 
        \raggedright Direct  & 
        Yes & Yes & -- & -- & Yes & Yes & Yes & 
        Yes & -- & Yes & -- & -- & -- & -- \\
\bottomrule
    \end{tabular}}
\end{table*}

\begin{table*}[htbp]
    \centering
    \scriptsize 
    \setlength{\tabcolsep}{2pt} 
    \renewcommand{\arraystretch}{1.3} 

    \caption{Payment Data Collection, Usage, and Disclosure Analysis}
    \label{tab:payment_data}
    \resizebox{\linewidth}{!}{
    \begin{tabular}{l | c | p{1cm} | *{7}{c} | *{7}{c} }
        \toprule
        \multirow{3}{*}{\textbf{Platform}} 
        & \multicolumn{16}{c}{\textbf{Payment Data}} \\
        \cmidrule(l){2-17}
        
        & \multirow{2}{*}{\rotatebox{90}{\textbf{Collect?}}} 
        & \multirow{2}{*}{\thead{Collect \\ Method} } 
        & \multicolumn{7}{c|}{\textbf{Use / Process}} 
        & \multicolumn{7}{c}{\textbf{Disclose / Share}} \\
        \cmidrule(lr){4-10} \cmidrule(l){11-17}

        & &  
        & \thead{Maintain \\ Service} 
        & \thead{User \\ Support} 
        & \thead{Improve / \\ Research} 
        & \thead{Communic- \\ ate} 
        & \thead{Security \\ Integrity} 
        & \thead{Legal \\ Purposes} 
        & \thead{Marketing} 
        & \thead{Service \\ Providers} 
        & \thead{Business \\ Transfers} 
        & \thead{Legal \\ Authorities} 
        & \thead{Related \\ Companies} 
        & \thead{Advertisers} 
        & \thead{Other 3rd \\ Parties} 
        & \thead{Public} \\
        \midrule

        \textbf{Grok (PP)} & 
        Yes & 
        \raggedright Through 3rd party & 
        Yes & -- & -- & -- & Yes & Yes & --& 
        Yes & Yes & Yes &-- & -- & -- & -- \\ 
        \textbf{Grok (PP-EU)} & 
        Yes & 

        \raggedright Through 3rd party & 
        Yes & -- & -- & -- & -- & -- & --& 
        Yes & -- & -- &-- & -- & -- & -- \\
\bottomrule
    \end{tabular}}
\end{table*}

\begin{table*}[htbp]
    \centering
    \scriptsize 
    \setlength{\tabcolsep}{2pt} 
    \renewcommand{\arraystretch}{1.3} 

    \caption{Communication Data Collection, Usage, and Disclosure Analysis}
    \label{tab:communication_data}
    \resizebox{\linewidth}{!}{
    \begin{tabular}{l | c | p{1cm} | *{7}{c} | *{7}{c} }
        \toprule
        \multirow{3}{*}{\textbf{Platform}} 
        & \multicolumn{16}{c}{\textbf{Communication Data}} \\
        \cmidrule(l){2-17}
        
        & \multirow{2}{*}{\rotatebox{90}{\textbf{Collect?}}} 
        & \multirow{2}{*}{\thead{Collect \\ Method} } 
        & \multicolumn{7}{c|}{\textbf{Use / Process}} 
        & \multicolumn{7}{c}{\textbf{Disclose / Share}} \\
        \cmidrule(lr){4-10} \cmidrule(l){11-17}

        & &  
        & \thead{Maintain \\ Service} 
        & \thead{User \\ Support} 
        & \thead{Improve / \\ Research} 
        & \thead{Communic- \\ ate} 
        & \thead{Security \\ Integrity} 
        & \thead{Legal \\ Purposes} 
        & \thead{Marketing} 
        & \thead{Service \\ Providers} 
        & \thead{Business \\ Transfers} 
        & \thead{Legal \\ Authorities} 
        & \thead{Related \\ Companies} 
        & \thead{Advertisers} 
        & \thead{Other 3rd \\ Parties} 
        & \thead{Public} \\
        \midrule

        \textbf{Grok (PP)} & 
        Yes & 
        \raggedright Direct & 
        Yes & Yes & Yes & -- & Yes & Yes & --& 
        Yes & Yes & Yes  & Yes & -- & -- & --\\ 
        \textbf{Grok (PP-EU)} & 
        Yes & 
        \raggedright Direct & 
        Yes & Yes & Yes & -- & Yes & Yes & --
         
        &Yes & -- & -- & -- & -- & -- & --  \\
\bottomrule
    \end{tabular}}
\end{table*}

\begin{table*}[htbp]
    \centering
    \scriptsize 
    \setlength{\tabcolsep}{2pt} 
    \renewcommand{\arraystretch}{1.3} 

    \caption{Interests and preferences Data Collection, Usage, and Disclosure Analysis}
    \label{tab:interest_data}
    \resizebox{\linewidth}{!}{
    \begin{tabular}{l | c | p{1cm} | *{7}{c} | *{7}{c} }
        \toprule
        \multirow{3}{*}{\textbf{Platform}} 
        & \multicolumn{16}{c}{\textbf{Interest and Preference Data}} \\
        \cmidrule(l){2-17}
        
        & \multirow{2}{*}{\rotatebox{90}{\textbf{Collect?}}} 
        & \multirow{2}{*}{\thead{Collect \\ Method} } 
        & \multicolumn{7}{c|}{\textbf{Use / Process}} 
        & \multicolumn{7}{c}{\textbf{Disclose / Share}} \\
        \cmidrule(lr){4-10} \cmidrule(l){11-17}

        & &  
        & \thead{Maintain \\ Service} 
        & \thead{User \\ Support} 
        & \thead{Improve / \\ Research} 
        & \thead{Communic- \\ ate} 
        & \thead{Security \\ Integrity} 
        & \thead{Legal \\ Purposes} 
        & \thead{Marketing} 
        & \thead{Service \\ Providers} 
        & \thead{Business \\ Transfers} 
        & \thead{Legal \\ Authorities} 
        & \thead{Related \\ Companies} 
        & \thead{Advertisers} 
        & \thead{Other 3rd \\ Parties} 
        & \thead{Public} \\
        \midrule

        \textbf{Grok (PP)} & 
        -- & 
        \raggedright -- & 
        -- & -- & -- & -- & -- & -- & --&
        -- & -- & -- &-- & -- & -- & -- \\  
        \textbf{Grok (PP-EU)} & 
        -- & 
        \raggedright -- & 
        -- & -- & -- & -- & -- & -- & --&
        -- & -- & -- &-- & -- & -- & -- \\
\bottomrule
    \end{tabular}}
\end{table*}

\begin{table*}[htbp]
    \centering
    \scriptsize 
    \setlength{\tabcolsep}{2pt} 
    \renewcommand{\arraystretch}{1.3} 

    \caption{Usage Data Collection, Usage, and Disclosure Analysis}
    \label{tab:usage_data}
    \resizebox{\linewidth}{!}{
    \begin{tabular}{l | c | p{1cm} | *{7}{c} | *{7}{c} }
        \toprule
        \multirow{3}{*}{\textbf{Platform}} 
        & \multicolumn{16}{c}{\textbf{Usage Data}} \\
        \cmidrule(l){2-17}
        
        & \multirow{2}{*}{\rotatebox{90}{\textbf{Collect?}}} 
        & \multirow{2}{*}{\thead{Collect \\ Method} } 
        & \multicolumn{7}{c|}{\textbf{Use / Process}} 
        & \multicolumn{7}{c}{\textbf{Disclose / Share}} \\
        \cmidrule(lr){4-10} \cmidrule(l){11-17}

        & &  
        & \thead{Maintain \\ Service} 
        & \thead{User \\ Support} 
        & \thead{Improve / \\ Research} 
        & \thead{Communic- \\ ate} 
        & \thead{Security \\ Integrity} 
        & \thead{Legal \\ Purposes} 
        & \thead{Marketing} 
        & \thead{Service \\ Providers} 
        & \thead{Business \\ Transfers} 
        & \thead{Legal \\ Authorities} 
        & \thead{Related \\ Companies} 
        & \thead{Advertisers} 
        & \thead{Other 3rd \\ Parties} 
        & \thead{Public} \\
        \midrule

        \textbf{Grok (PP)} & 
        Yes & 
        \raggedright Direct & 
        Yes & Yes & Yes & -- & Yes & Yes & No&
        Yes & Yes & Yes & Yes & No & --  &--\\  
        \textbf{Grok (PP-EU)} & 
        Yes & 
        \raggedright Direct & 
        Yes & Yes & -- & -- & Yes & Yes & Yes&
        Yes & -- & -- & -- & -- & --  &--\\
\bottomrule
    \end{tabular}}
\end{table*}

\begin{table*}[htbp]
    \centering
    \scriptsize 
    \setlength{\tabcolsep}{2pt} 
    \renewcommand{\arraystretch}{1.3} 

    \caption{Feedback Data Collection, Usage, and Disclosure Analysis}
    \label{tab:feedback_data}
    \resizebox{\linewidth}{!}{
    \begin{tabular}{l | c | p{1cm} | *{7}{c} | *{7}{c} }
        \toprule
        \multirow{3}{*}{\textbf{Platform}} 
        & \multicolumn{16}{c}{\textbf{Feedback Data}} \\
        \cmidrule(l){2-17}
        
        & \multirow{2}{*}{\rotatebox{90}{\textbf{Collect?}}} 
        & \multirow{2}{*}{\thead{Collect \\ Method} } 
        & \multicolumn{7}{c|}{\textbf{Use / Process}} 
        & \multicolumn{7}{c}{\textbf{Disclose / Share}} \\
        \cmidrule(lr){4-10} \cmidrule(l){11-17}

        & &  
        & \thead{Maintain \\ Service} 
        & \thead{User \\ Support} 
        & \thead{Improve / \\ Research} 
        & \thead{Communic- \\ ate} 
        & \thead{Security \\ Integrity} 
        & \thead{Legal \\ Purposes} 
        & \thead{Marketing} 
        & \thead{Service \\ Providers} 
        & \thead{Business \\ Transfers} 
        & \thead{Legal \\ Authorities} 
        & \thead{Related \\ Companies} 
        & \thead{Advertisers} 
        & \thead{Other 3rd \\ Parties} 
        & \thead{Public} \\
        \midrule

        \textbf{Grok (PP)} & 
        Yes & 
        \raggedright Direct & 
        Yes & Yes & Yes & No & Yes & Yes & --&
        Yes & Yes & Yes & Yes & -- & No  &--\\  
        \textbf{Grok (PP-EU)} & 
        Yes & 
        \raggedright Direct & 
        Yes & Yes & Yes & -- & Yes & Yes & Yes&
        Yes & -- & -- & -- & -- & --  &--\\
\bottomrule
    \end{tabular}}
\end{table*}

\begin{table*}[htbp]
    \centering
    \scriptsize 
    \setlength{\tabcolsep}{2pt} 
    \renewcommand{\arraystretch}{1.3} 

    \caption{Social Media Data Collection, Usage, and Disclosure Analysis}
    \label{tab:socialmedia_data}
    \resizebox{\linewidth}{!}{
    \begin{tabular}{l | c | p{1cm} | *{7}{c} | *{7}{c} }
        \toprule
        \multirow{3}{*}{\textbf{Platform}} 
        & \multicolumn{16}{c}{\textbf{Social Media Data}} \\
        \cmidrule(l){2-17}
        
        & \multirow{2}{*}{\rotatebox{90}{\textbf{Collect?}}} 
        & \multirow{2}{*}{\thead{Collect \\ Method} } 
        & \multicolumn{7}{c|}{\textbf{Use / Process}} 
        & \multicolumn{7}{c}{\textbf{Disclose / Share}} \\
        \cmidrule(lr){4-10} \cmidrule(l){11-17}

        & &  
        & \thead{Maintain \\ Service} 
        & \thead{User \\ Support} 
        & \thead{Improve / \\ Research} 
        & \thead{Communic- \\ ate} 
        & \thead{Security \\ Integrity} 
        & \thead{Legal \\ Purposes} 
        & \thead{Marketing} 
        & \thead{Service \\ Providers} 
        & \thead{Business \\ Transfers} 
        & \thead{Legal \\ Authorities} 
        & \thead{Related \\ Companies} 
        & \thead{Advertisers} 
        & \thead{Other 3rd \\ Parties} 
        & \thead{Public} \\
        \midrule

        \textbf{Grok (PP)} & 
        Yes & 
        \raggedright Through 3rd party & 
        Yes & Yes & Yes & No & Yes & Yes & --&
        Yes & Yes & Yes & Yes & -- & Yes  &--\\  
        \textbf{Grok (PP-EU)} & 
        Yes & 
        \raggedright Through 3rd party & 
        -- & -- & -- & -- & -- & -- & Yes&
        -- & -- & -- & -- & Yes & --  &--\\
\bottomrule
    \end{tabular}}
\end{table*}

\begin{table*}[htbp]
    \centering
    \scriptsize 
    \setlength{\tabcolsep}{2pt} 
    \renewcommand{\arraystretch}{1.3} 

    \caption{Technical Data Collection, Usage, and Disclosure Analysis}
    \label{tab:technical_data}
    \resizebox{\linewidth}{!}{
    \begin{tabular}{l | c | p{1cm} | *{7}{c} | *{7}{c} }
        \toprule
        \multirow{3}{*}{\textbf{Platform}} 
        & \multicolumn{16}{c}{\textbf{Technical Data}} \\
        \cmidrule(l){2-17}
        
        & \multirow{2}{*}{\rotatebox{90}{\textbf{Collect?}}} 
        & \multirow{2}{*}{\thead{Collect \\ Method} } 
        & \multicolumn{7}{c|}{\textbf{Use / Process}} 
        & \multicolumn{7}{c}{\textbf{Disclose / Share}} \\
        \cmidrule(lr){4-10} \cmidrule(l){11-17}

        & &  
        & \thead{Maintain \\ Service} 
        & \thead{User \\ Support} 
        & \thead{Improve / \\ Research} 
        & \thead{Communic- \\ ate} 
        & \thead{Security \\ Integrity} 
        & \thead{Legal \\ Purposes} 
        & \thead{Marketing} 
        & \thead{Service \\ Providers} 
        & \thead{Business \\ Transfers} 
        & \thead{Legal \\ Authorities} 
        & \thead{Related \\ Companies} 
        & \thead{Advertisers} 
        & \thead{Other 3rd \\ Parties} 
        & \thead{Public} \\
        \midrule

        \textbf{Grok (PP)} & 
        Yes & 
        \raggedright Direct & 
        Yes & Yes & Yes & Yes & Yes & Yes & --&
        Yes & Yes & Yes & Yes & -- & Yes  &--\\  
        \textbf{Grok (PP-EU)} & 
        Yes & 
        \raggedright Direct & 
        Yes & Yes & Yes & -- & Yes & Yes & Yes&
        Yes & -- & -- & -- & -- & --  &--\\
\bottomrule
    \end{tabular}}
\end{table*}

\begin{table*}[htbp]
    \centering
    \scriptsize 
    \setlength{\tabcolsep}{2pt} 
    \renewcommand{\arraystretch}{1.3} 

    \caption{Publicly Available Data Collection, Usage, and Disclosure Analysis}
    \label{tab:public_data}
    \resizebox{\linewidth}{!}{
    \begin{tabular}{l | c | p{1cm} | *{7}{c} | *{7}{c} }
        \toprule
        \multirow{3}{*}{\textbf{Platform}} 
        & \multicolumn{16}{c}{\textbf{Publicly Available Data}} \\
        \cmidrule(l){2-17}
        
        & \multirow{2}{*}{\rotatebox{90}{\textbf{Collect?}}} 
        & \multirow{2}{*}{\thead{Collect \\ Method} } 
        & \multicolumn{7}{c|}{\textbf{Use / Process}} 
        & \multicolumn{7}{c}{\textbf{Disclose / Share}} \\
        \cmidrule(lr){4-10} \cmidrule(l){11-17}

        & &  
        & \thead{Maintain \\ Service} 
        & \thead{User \\ Support} 
        & \thead{Improve / \\ Research} 
        & \thead{Communic- \\ ate} 
        & \thead{Security \\ Integrity} 
        & \thead{Legal \\ Purposes} 
        & \thead{Marketing} 
        & \thead{Service \\ Providers} 
        & \thead{Business \\ Transfers} 
        & \thead{Legal \\ Authorities} 
        & \thead{Related \\ Companies} 
        & \thead{Advertisers} 
        & \thead{Other 3rd \\ Parties} 
        & \thead{Public} \\
        \midrule

        \textbf{Grok (PP)} & 
        Yes & 
        \raggedright Through 3rd party & 
        Yes & -- & Yes & -- & Yes & Yes & --&
        Yes & Yes & Yes & Yes & -- & --  &--\\  
        \textbf{Grok (PP-EU)} & 
        Yes & 
        \raggedright  Through 3rd party& 
        -- & -- & Yes & -- & -- & -- & --&
        -- & -- & -- & -- & -- & --  &--\\
\bottomrule
    \end{tabular}}
\end{table*}

\begin{table*}[htbp]
    \centering
    \scriptsize 
    \setlength{\tabcolsep}{2pt} 
    \renewcommand{\arraystretch}{1.3} 

    \caption{Face and Head Movement Data Collection, Usage, and Disclosure Analysis}
    \label{tab:face_data}
    \resizebox{\linewidth}{!}{
    \begin{tabular}{l | c | p{1cm} | *{7}{c} | *{7}{c} }
        \toprule
        \multirow{3}{*}{\textbf{Platform}} 
        & \multicolumn{16}{c}{\textbf{Face and Head Movement Data}} \\
        \cmidrule(l){2-17}
        
        & \multirow{2}{*}{\rotatebox{90}{\textbf{Collect?}}} 
        & \multirow{2}{*}{\thead{Collect \\ Method} } 
        & \multicolumn{7}{c|}{\textbf{Use / Process}} 
        & \multicolumn{7}{c}{\textbf{Disclose / Share}} \\
        \cmidrule(lr){4-10} \cmidrule(l){11-17}

        & &  
        & \thead{Maintain \\ Service} 
        & \thead{User \\ Support} 
        & \thead{Improve / \\ Research} 
        & \thead{Communic- \\ ate} 
        & \thead{Security \\ Integrity} 
        & \thead{Legal \\ Purposes} 
        & \thead{Marketing} 
        & \thead{Service \\ Providers} 
        & \thead{Business \\ Transfers} 
        & \thead{Legal \\ Authorities} 
        & \thead{Related \\ Companies} 
        & \thead{Advertisers} 
        & \thead{Other 3rd \\ Parties} 
        & \thead{Public} \\
        \midrule

        \textbf{Grok (PP)} & 
        -- & 
        \raggedright -- & 
        -- & -- & -- & -- & -- & -- & --&
        -- & -- & -- & -- & -- & --  &--\\  
        \textbf{Grok (PP-EU)} & 
        -- & 
        \raggedright -- & 
        -- & -- & -- & -- & -- & -- & --&
        -- & -- & -- & -- & -- & --  &--\\
\bottomrule
    \end{tabular}}
\end{table*}



\end{document}